\documentclass[conference, 
               ]{IEEEtran}
\pdfoutput=1
\IEEEoverridecommandlockouts
\usepackage{cite}
\usepackage{amsmath,amssymb,amsfonts}
\usepackage{algorithmic}
\usepackage{graphicx}
\usepackage{textcomp}
\usepackage{tabularx}
\usepackage{xcolor}
\def\BibTeX{{\rm B\kern-.05em{\sc i\kern-.025em b}\kern-.08em
    T\kern-.1667em\lower.7ex\hbox{E}\kern-.125emX}}

\usepackage[hidelinks]{hyperref}
\usepackage[nolist]{acronym}
\usepackage{siunitx}
\usepackage{booktabs}
\usepackage{graphicx}
\usepackage[caption=false, font=footnotesize]{subfig}

\begin{document}

\title{Electricity Demand Forecasting in Future Grid States: A Digital Twin-Based Simulation Study}

\makeatletter
\newcommand{\linebreakand}{%
  \end{@IEEEauthorhalign}
  \hfill\mbox{}\par
  \mbox{}\hfill\begin{@IEEEauthorhalign}
}
\makeatother

\author{
	\IEEEauthorblockN{Daniel R. Bayer}
	\IEEEauthorblockA{\textit{Modeling and Simulation} \\
	\textit{University of W\"urzburg}\\
	W\"urzburg, Germany \\
	daniel.bayer@uni-wuerzburg.de}
	\and
	\IEEEauthorblockN{Felix Haag}
	\IEEEauthorblockA{\textit{Information Systems and Energy Efficient Systems} \\
	\textit{University of Bamberg}\\
	Bamberg, Germany \\
	felix.haag@uni-bamberg.de}
	\and
	\IEEEauthorblockN{Marco Pruckner}
	\IEEEauthorblockA{\textit{Modeling and Simulation} \\
	\textit{University of W\"urzburg}\\
	W\"urzburg, Germany \\
	marco.pruckner@uni-wuerzburg.de}
 \linebreakand
	\IEEEauthorblockN{Konstantin Hopf}
	\IEEEauthorblockA{\textit{Information Systems and Energy Efficient Systems} \\
	\textit{University of Bamberg}\\
	Bamberg, Germany \\
	konstantin.hopf@uni-bamberg.de}
}

%\author{
%    \IEEEauthorblockN{Daniel R. Bayer\IEEEauthorrefmark{1},
%                      Felix Haag\IEEEauthorrefmark{2},
%                      Marco Pruckner\IEEEauthorrefmark{1} and
%                      Konstantin Hopf\IEEEauthorrefmark{2}}
%    \IEEEauthorblockA{\IEEEauthorrefmark{1}Modeling and Simulation, University of W\"urzburg, 97074 W\"urzburg, Germany\\
%Email: \{daniel.bayer, marco.pruckner\}@uni-wuerzburg.de}
%\IEEEauthorblockA{\IEEEauthorrefmark{2}Information Systems and Energy Efficient Systems, University of Bamberg, 96047 Bamberg, Germany\\
%Email: \{felix.haag, konstantin.hopf\}@uni-bamberg.de}
%}

\maketitle
%\IEEEpeerreviewmaketitle

\begin{abstract}
Short-term forecasting of residential electricity demand is an important task for utilities. Yet, many small and medium-sized utilities still use simple forecasting approaches such as \acfp{SLP}, which treat residential households similarly and neither account for renewable energy installations nor novel large consumers (e.g., heat pumps, electric vehicles). The effectiveness of such ``one-fits-all'' approaches in future grid states---where decentral generation and sector coupling increases---are questionable. Our study challenges these forecasting practices and investigates whether \acf{ML} approaches are suited to predict electricity demand in today's and in future grid states. We use real smart meter data from 3{,}511 households in Germany over 34 months. We extrapolate this data with future grid states (i.e., increased decentral generation and storage) based on a digital twin of a local energy system.
%To obtain proper future system scenarios, we regionalize the national photovoltaics expansion targets.
Our results show that \acl{LSTM} approaches outperform \acp{SLP} as well as simple benchmark estimators with up to 68.5\% lower \acl{RMSE} for a day-ahead forecast, especially in future grid states. Nevertheless, all prediction approaches perform worse in future grid states. Our findings therefore reinforce the need (a) for utilities and grid operators to employ \ac{ML} approaches instead of traditional demand prediction methods in future grid states and (b) to prepare current \ac{ML} methods for future grid states.
%Our future study will significantly extend our predictive and simulative modeling. %In the end, we aim to inform the forecasting research and practice about potential operational and monetary improvements that can be realized with effective residential demand forecasting, today and in the future.
\end{abstract}

\begin{IEEEkeywords}
Digital twin, Demand forecasting, Future energy systems, Simulation, Neural networks
\end{IEEEkeywords}

%\section*{List of Acronyms}
\begin{acronym}[ABCDEF]
	%\acro{ANN}{Artificial Neural Network}
    \acro{ARIMA}{Autoregressive Integrated Moving Average}
    \acro{BM}{Baseline model}
	\acro{CNN}{Convolutional Neural Network}
	\acro{CS}{Current State}
	\acro{LSTM}{Long Short-Term Memory}
	\acro{ML}{Machine Learning}
	\acro{PV}{Photovoltaics}
	\acro{MAE}{Mean Absolute Error}
	\acro{MSE}{Mean Square Error}
	\acro{RES}{Renewable Energy Sources}
	\acro{RNN}{Recurrent Neural Network}
	%\acro{RQ}{Research Question}
	\acro{RMSE}{Root Mean Square Error}
	\acro{MAPE}{Mean Absolute Percentage Error}
	\acro{SLP}{Synthesized Load Profile}
	\acro{BESS}{Battery Energy Storage System}
	\acro{HP}{Heat Pump}
\end{acronym}

\textit{This is the author's version of the work. It is posted here for your personal use, not for redistribution. Please cite the officially published version in the Proceedings of the 9th International Conference on Smart and Sustainable Technologies (SpliTech 2024), June 25--28, 2024, Bol and Split, Croatia \url{https://doi.org/10.23919/SpliTech61897.2024.10612563}.}

\textit{\copyright 2024 IEEE. Personal use of this material is permitted.  Permission from IEEE must be obtained for all other uses, in any current or future media, including reprinting/republishing this material for advertising or promotional purposes, creating new collective works, for resale or redistribution to servers or lists, or reuse of any copyrighted component of this work in other works.}

\section{Introduction}

\acresetall
Residential consumers play an important role when it comes to operating the electricity grid.
%Around a quarter of total household energy demand can be directly attributed to electricity use  \cite{eurostat_EnergyConsumptionHouseholds_2020}.
While private households' electric grid demand has been stable for a long time, the energy transition towards renewable and distributed generation considerably changes residential demand profiles. 
Additional loads resulting from sector coupling (e.g., heat pumps or electric vehicles)\cite{2020_Doluweera_EVImpactOnGridAndSustainabilityInAlberta} and local energy generation technologies (e.g., rooftop \ac{PV} and battery storage systems) \cite{2019_Blasi_ImpactsRenewablesBESS_DSO} alter demand profiles. Both developments change the individual electricity demand pattern strongly \cite{2023_Bayer_DataDrivenDemandAndSPFModelingForHPs} and complicate electricity forecasting \cite{wang_Review_Smart_2019}. Our paper  focuses on the changed grid demand pattern of individual buildings that is caused by adding \ac{PV} and battery installations. We thereby do not consider the surplus local \ac{PV} power that is fed into the grid.

Two technical developments make electric short-term demand forecasting viable, even for smaller utilities and grid operators. First, the steady expansion of the smart meter infrastructure, which can measure demand and production data at the level of individual households on 15 minute to one hour time intervals, and communicates this data to the grid operator. The new data on \textit{individual households} enable more precise modeling of residential electricity use (or production).
%We use this smart meter data as basis for modeling a digital twin of the local energy system.
Second, innovations in electric demand forecasting methods have significantly reduced errors in short- and long-term time intervals (as narrative \cite{haben_review_2021} and quantitative reviews \cite{hopf_meta-regression_2023} show).
Recent improvements in forecasting quality are driven by applying \ac{ML} methods, like artificial neural networks. Currently, one of the most frequently used neural network architectures in time-series forecasting research is \ac{LSTM}\cite{haben_review_2021}. % outperforming earlier methods \cite{kim_predicting_2019, kong_short_term_2019}. %, alhussein_hybrid_2020}.
%\ac{ML} methods, such as \acp{ANN}, have led to recent improvements in forecasting quality. Long Short-Term Memory (LSTM) is a commonly used ANNs architecture in time-series forecasting research due to its superior forecasting performance compared to earlier methods.
However, current research focuses on investigating the performance of deep-learning approaches for demand forecasting using actual grid data. Yet, it remains unexplored how well these approaches perform in future grid states showing increased decentral production and additional loads, e.g., through heat pumps.

%With the expansion of the smart metering infrastructure and the increased availability of the data collected from these meters researchers can create a digital twin of a complete local energy system \cite{2023_Bayer_DT_LocalEnergySystem}. Such a digital twin aims to mimic the addition of new \ac{PV} installations or heat pumps as realistically as possible based on today's demand profiles and a predefined expansion scenario. As an output we get the modified individual energy grid demand. The demand sum of all households can directly be used as prediction input.

Despite these technical developments, many utility companies in practice still rely on simple forecasting approaches like standardized \acp{SLP}, which generalize all households to a single averaged load profile. %, as the expansion of smart meters, including the communication infrastructure, is only just beginning in many areas.
%They do so, because smart meter data is not available for all of their households (e.g., due to delayed roll-out of smart meters) or because the most efficient methods have not found application in productive systems, given that the business need to employ more efficient forecasting methods is not urgent enough.
Reasons for this are the missing business need for more effective forecasts or the lack of availability of the most efficient forecasting methods in productive systems. We argue that \acp{SLP} and other simple estimators may be effective today, but their performance is questionable in the future, as these approaches cannot account for weather-dependent, highly fluctuating loads like heat pumps and renewable generation such as \ac{PV}. Thus, this study investigates the research question: \textit{To what extent can ML approaches (i.e., \ac{LSTM}-based models) predict the short-term electricity demand of residential households at the grid level in a (a) current and (b) future grid setting, where decentral generation and large consumers are more prominent?}

To do so, we use an already existing and validated digital twin of a complete local energy system \cite{2021_CityDTPotentials_Review,2020_SmartCityDT_RealTimeBuildingEnergyBenchmarking,2018_SmartCityDigitalTwin}. Researchers and utility companies can leverage increased smart meter data availability with such digital twins, which mimic the addition of new \ac{PV} installations, heat pumps, etc. as realistically as possible based on today's demand profiles and a predefined expansion scenario \cite{2023_Bayer_DT_LocalEnergySystem}. With the digital twin, we obtain the modified individual energy grid demand for future grid states and use this data as prediction input.
We instantiate \ac{LSTM} prediction models and use a \ac{SLP} to capture the current state of practice in many utility companies for short-term demand forecasting. In addition, we use a real dataset from a regional electricity provider in Germany and analyze the prediction performance using data on the \ac{CS} of the electricity grid (see research question a) and several simulated future grid states (see research question b). The novelty of our study lies in the combination of predictive modeling approaches and a digital twin of a local energy system to examine the performance of contemporary forecasting methods in future grid states.

%After reviewing related work in the area of short-term electric demand forecasting in the next section, we describe our research approach and the interim findings. We close with a discussion of the results, their implications, and we outline our future work.

\section{Related work}% in short-term electricity demand forecasting}

%\begin{itemize}
%  \item short-term
%  \item Comparison with synthesized load profiles
%  \item Probablistic forecasting -> vom Scheidt (2021, e-Energy)
%  \item Forecasting at different levels: grid, household level
%  \item Which data? vom Scheidt (2021) use simulated data, we have real data
%  \item	\ac{LSTM} + Transformers kurz erklären
%\end{itemize}

Literature suggests several approaches to predict the short-term electricity demand of households \cite{hong_probabilistic_2016}. Besides statistical methods, \ac{ML} gains popularity because these methods usually offer a higher predictive performance \cite{haben_review_2021, vom_scheidt_probabilistic_2021, arpogaus_probabilistic_2021}. Examples of commonly used \ac{ML} approaches include tree- \cite{liao_research_2019}, kernel- \cite{sousa_short_term_2013}, and nearest neighbor-based \cite{desilva_incremental_2011} methods. 
In addition, neural networks have attracted much attention and led to numerous applications in load forecasting \cite{haben_review_2021}. Simple approaches include feed-forward networks or the multi-layer perceptron, which researchers frequently call ``vanilla'' methods, in particular when they comprise only a single hidden layer. % \cite{hastie_statistical_learning_2001}. 
Despite the simplicity of vanilla methods, they often have a high predictive performance compared to conventional \ac{ML} methods \cite{moon_comparative_2019}. 

In the last decade, ``deep learning'' architectures with numerous hidden layers gained attraction. Most popular among these are \acp{LSTM} \cite{haben_review_2021}. %, which include various gates that select which parts of the data input are important (input-gate), decide when past data inputs are ``forgotten'' (forget-gate), and what is relevant to the output at each time step (output-gate) \cite{hochreiter_long_1997}. 
Although there are other deep learning architectures in the demand forecasting field with promising performance results (e.g., \acp{CNN} \cite{kim_predicting_2019, voss_residential_2018} and the transformer architecture \cite{giacomazzi_2023}), the \ac{LSTM} seems to have a comparably high predictive performance at a manageable computational effort \cite{giacomazzi_2023,vaswani_attention_2017}, which is a relevant criterion for our simulation study.
Recent publications propose \cite{2022_DTOverviewInEnergyGrids} the use of digital twins as a data generator to train prediction algorithms.
For example, \cite{2022_Henzel_BuildingEnergyConsumptionForecasting_DT} and \cite{2022_DTIndividualShortTermEnergyPrediction} use a digital twin of a single, individual building for day-ahead energy consumption forecasting.
In \cite{2022_DTBasedDayAheadEnergySystemScheduling}, a digital twin is used for day-ahead scheduling in a local energy system with a high share of renewable energy systems.
%To our best knowledge, a digital twin of a local energy system has not yet been used for demand prediction on the grid- or utility-level.
In our review of related works, we have not found studies that examine state-of-the-art deep-learning approaches for load forecasting in \textit{future} grid states, which requires simulated data, e.g., generated by a digital twin.

\section{Research approach}

We compute the forecasting performance for two cases (see Fig.~\ref{fig:exp_design}) to answer the research question using \ac{LSTM}- and \ac{SLP}-based approaches to examine the current state of research and practice.

\begin{figure}[t]
\centering
\includegraphics[width=\linewidth]{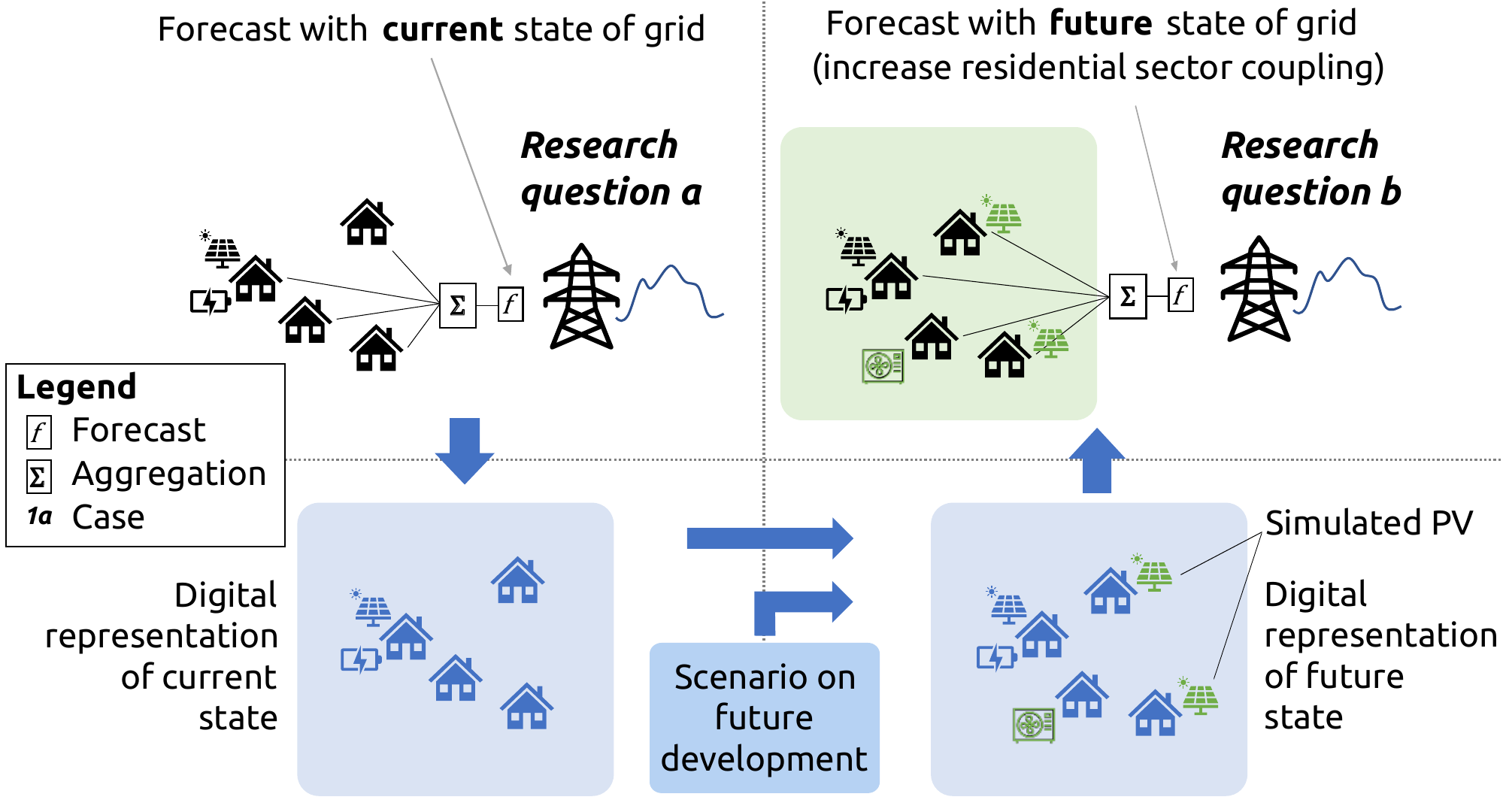}
\caption{Experimental design.}\label{fig:exp_design}
%\vspace*{-0.5em}
\end{figure}
In the following, we describe the simulation approach for obtaining future grid states, the dataset that represents the current grid state, the derivation of the future grid state based on a regionalization of the German goals for \ac{PV} expansion, and our predictive modeling.

\subsection{Digital twin modeling and simulation of future grid states}
To obtain demand profiles for grid states that model the future as realistically as possible, we employ a previously developed digital twin of the existing energy system of Ha{\ss}furt, a town in Southern Germany \cite{2023_Bayer_DT_LocalEnergySystem}.
The digital twin aims to mimic the behavior of the complete electric energy system (i.e., with increased \ac{PV}, \ac{PV}-battery-combination, and \ac{HP} penetration rates) based on today's demand profiles, with a particular focus on residential buildings. Inside the digital twin, every residential building is represented individually based on the recorded smart meter data. If a building is selected for adding a \ac{PV} installation (and possibly a battery storage system), we use geospatial data to determine the orientation and maximal installable rooftop \ac{PV} power. These added components can alter the building energy demand per time step. If a battery is added, we use a rule-based control strategy to maximize local \ac{PV} self-consumption. Such a rule-based strategy seems sensible, as increasing self-consumption is one of the main arguments for installing a battery in a residential building, especially in Germany \cite{2020_Figgener_BESS_GermanMarketReview}.
Every added battery has a power of 7~kW and a capacity of \num{10.5}~kWh. The choice of this power and power-to-energy ratio seems appropriate, as it represents the median of currently installed residential batteries in Germany \cite{2020_Figgener_BESS_GermanMarketReview}.

\subsection{Dataset on current grid state}
Besides the information on the structure of the energy system of the considered town, we acquired the smart meter data containing demand and feed-in time series for all connected households of the town (see ``Actual data'' in TABLE~\ref{tab:data_distribution}).
We focus on the recorded time series of the residential households between 01/2019 and 12/2021 with an hourly resolution.
After removing erroneous and incomplete time series and those that could not be attributed to an existing address, we have a remaining number of 3{,}511 households (counting multi-story buildings as one household), capturing almost all residential buildings in the considered town.
The profiles for simulating new rooftop \ac{PV} installations are taken from existing, individually metered ones.

\subsection{PV expansion scenarios for modeling future grid states}
The paper aims to simulate the expected system state in the year 2037.
To obtain a reasonable scenario for 2037, we regionalize the national expansion targets for Germany as defined by \cite{2022_Szenariorahmen_NEP} according to a statistical feature available on both the country and the local level.
Therefore, we use the number of residential buildings, as it is the best feature available for explaining the current state (in terms of \ac{PV} and battery power in our considered town).
%(estimation using this feature: 7.0~MW installed \ac{PV} power, actual value is 9.0~MW in 2021).
In 2021, the total power of rooftop \ac{PV} systems was \num{9.0}~MW and the total power of residential battery storage systems was \num{425}~kW in our town.
Using the above method, we regionalize the country-wide targets for 2037 \cite{2022_Szenariorahmen_NEP} of \num{345.4}~GW \ac{PV} power and \num{67.4}~GW residential battery power to a local value of \num{32}~MW rooftop \ac{PV} power and \num{15.2}~MW residential battery power.
We define this scenario as \textit{S1}.
To test the prediction algorithms on different scenarios, we define a second scenario S2, in which we extrapolate the development to date on a linear basis. In this scenario, we have an addition of 19 MW of rooftop \ac{PV} power and 8.8 MW of residential battery power until 2037.

\subsection{Simulation results for current and future grid states}
%\subsubsection*{Overview of expanded buildings}
We virtually equip 94\% of the buildings that currently do not have a \ac{PV} system with such a system in S1. In S2, this number reduces to 62\%. The number of buildings, where a battery storage system is added to the simulated \ac{PV} installation is set accordingly to reach the expansion goals as defined above. The total number of buildings with simulatively added components is presented in TABLE~\ref{tab:data_distribution}.
Nevertheless, we initially select the households for which we simulated \ac{PV} and battery storage systems randomly.
To ensure that the resulting grid demand is still representative, the simulation is executed multiple times.

\begin{table}[h]
\caption{Building type distribution in the dataset and the simulated scenarios.}
\label{tab:data_distribution}
\centering
\begin{tabularx}{\linewidth}{Xrcrr}
	\toprule
	&Actual data & & \multicolumn{2}{c}{Simulated data}\\
	\cmidrule{2-2}
	\cmidrule{4-5}
	& & & S1 & S2\\
	\midrule
	Residential buildings            & 3511 & & 3511 & 3511 \\
	... without \acs{PV} / PV+battery   & 3017 & &  189 & 1328 \\
	... with \acs{PV} (no battery)      &  377 & & 1179 &  888 \\
	... with \acs{PV} + battery         &  117 & & 2143 & 1295 \\
	%        ... with heat pump              & 114 & 114 & 114 \\
	%        ... with electric vehicle       &   ? &   ? &   ? \\
	\bottomrule
\end{tabularx}
\end{table}

To describe the \acf{CS} and the simulated electricity demand data, we compare three scenarios (named CS, S1, S2 respectively). We present the grid load caused by residential buildings (including negative values due to \ac{PV} grid feed-in) and the sum of the residential demand (i.e., the feed-in of surplus \ac{PV} power at building level is not taken into account) of all three scenarios for one example year in Fig.~\ref{fig:boxplot_accum_load}. As an individual \ac{PV} installation can outrage the local consumption, a negative load can occur on grid level. The negative load is not accumulated to compute the residential demand. For the latter prediction, we use the computed time series of the residential demand.
The comparison shows that the \ac{CS} has a less volatile grid load and demand than the future scenarios (S1 and S2). The median energy demand in the \ac{CS} setting is \num{1613}~kW and decreases respectively to \num{816}~kW (S1) and \num{1102}~kW (S2), while the standard deviation increases compared to the \ac{CS} by a factor of \num{3.8} for S1 and \num{6.0} for S2.
This comes along with the fact that we see a feed-in into upper grid levels in 26-31\% of all hours, which is not present in the baseline scenario.

\begin{figure}[h]
\centering
\includegraphics[width=\linewidth]{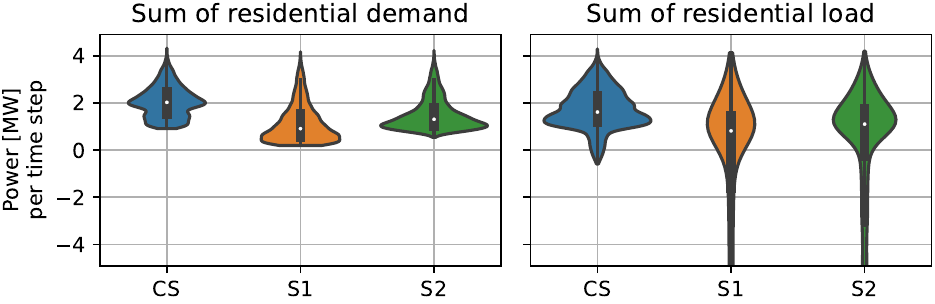}
\caption{Left: Violinplot of one-year electric demand on grid level caused by residential buildings in kW averaged over one hour for current grid state (CS) and the scenarios S1 and S2. Right: Same plot for electric load in kW. The negative values occur due to a strong feed-in from residential PV installations that cannot be consumed locally.}
\label{fig:boxplot_accum_load}
\end{figure}

\subsection{Predictive modeling}

%We divide the predictive modeling for estimating electricity demand in the current and future grid states into three steps.

We evaluate several \ac{ML} methods with the datasets for the current and future grid states following a univariate multistep time series forecasting setup, capturing the current state-of-the-art in \ac{ML}-based electricity demand forecasting. We choose to forecast the day-ahead sum of residential demand (see the left plot in Fig.~\ref{fig:boxplot_accum_load}) in an hourly resolution (i.e., the following 24 time steps). This choice aligns with the granularity of our dataset and is consistent with prevailing practices in demand forecasting \cite{haben_review_2021}. 
%We present the results for the demand prediction, i.e., the time series includes the feed-in of surplus \ac{PV} power and therefore includes negative values, as described above.
%
Our analysis starts with a vanilla \ac{LSTM}, similar to \cite{masum_MultistepTimeSeries_2018}. We also consider a more sophisticated architecture that combines an \ac{LSTM} with a \ac{CNN} as part of a Decoder-Encoder architecture \cite{kim_predicting_2019}.
Our implementation applies the \ac{MSE} loss function, the ``Adam'' algorithm as an optimizer, 200 training epochs for the \ac{LSTM} and 50 epochs for the \ac{CNN} as the latter tend to overfit\footnote{\ac{LSTM} parameters: units/\ac{LSTM}=100, activation=TanH; CNN-LSTM parameters: units/\ac{LSTM}=200, filters/Conv1D=32, kernel\textunderscore size/Conv1D=3, pool\textunderscore size/MaxPooling1D=2, units/ Dense=100, activation: LeakyReLu.}.
%Here, the encoder reads and represents the input sequence by learning from the patterns in the time series data. The encoder comprises two \ac{CNN} layers and a subsequent max-pooling layer. The decoder, consisting of an \ac{LSTM} and a fully connected layer, interprets the sequence and leverages it for prediction. 

\subsubsection{Training and data preparation}

For training the \ac{ML} models, we use data from the first 664 days of the \ac{CS} (case \textit{a}) and future grid states (case \textit{b}). To validate all estimators including the calculation of the final error metrics, we use the remaining 365 days.
%To enable a realistic validation scenario for practice, we train our models once on historical data and rely for testing on walk-forward validation.
This is an out-of-sample back-testing approach to evaluate how well a model makes a prediction for each time step (i.e., one day with 24 hours).
%This approach leverages a prior time step $[t_{x-1},t_x]$ to predict $t_{x+1}$. For validation, we compute the prediction error for $t_{x+1}$ using with the actual test data.
%Hence, we implement our model validation as a sliding window that emulates realistic prediction scenarios by incorporating prior data in each prediction step.

\subsubsection{Evaluation metrics}
We choose \ac{RMSE} and \ac{MAPE} as error metrics, which are commonly used in energy forecasting \cite{haben_review_2021}. The \ac{MAPE} is an extension of the \ac{MAE} and indicates the average absolute percentage deviation of the predicted value from the actual value.
%It is formally defined as
%\[
%\ac{MAPE} = \frac{100\%}{n} \sum_{i=1}^{n} \biggl| \frac{\hat{y}_{i} - y_{i}}{y_{i}} \biggl|,
%\]
%
%where $y_{i}$ describes the actual target value for the aggregated measurement observation $i$ (i.e., total demand for each hour) with $i\in\{1,\ldots,n\}$, while $\hat{y}_{i}$ denotes the model's prediction.
Although energy forecasting studies frequently use \ac{MAPE} \cite{haben_short_2019,kim_predicting_2019}, it has some drawbacks. The \ac{MAPE} is undefined if the ground truth equals zero, and produces a large percentage error when the ground truth values are small. Furthermore, \ac{MAPE} is systematically biased in favor of underestimates \cite{kim_NewMetricAbsolute_2016, shcherbakov_SurveyForecastError_2013}.

The \ac{RMSE} describes the spread of the residuals in the scale of the target variable \cite{kim_predicting_2019, masum_MultistepTimeSeries_2018, zhang_power_2021}.
%and is defined as
%\[
%\ac{RMSE} = \sqrt{\frac{1}{n} \sum_{i=1}^{n} (\hat{y}_{i} - y_{i})^{2}}.
%\]
%
In contrast to other metrics such as \ac{MAPE} and \ac{MAE}, \ac{RMSE} squares the residuals, giving a large weight to high deviations from the actual value and thus penalizing high errors. This characteristic is desirable for our case, as high prediction errors may involve high costs for utilities.

\subsubsection{Benchmark predictors}

Three benchmark predictors help us to interpret the performance of the \ac{ML} models. We include \acp{SLP} as a widespread method for residential demand forecasting in practice (for current grid state only, as there are no profiles for future grid state available). In addition, we first use naive estimators, incorporating ``Day before'' (i.e., the demand values of the previous day) and ``Day one week ago'' (i.e., the demand values of the seven days ago).
Finally, we train an \ac{ARIMA} model on the past five days for predicting day-ahead demand.

\section{Results}

We list the predictive performance of all estimators in TABLE~\ref{tab:overall-rmse-results} (overall \ac{RMSE} and \ac{MAPE}) and in Fig.~\ref{fig:forecast_results} (\ac{RMSE} unfolded over all hours of the day), for the test data for the \ac{CS} and future grid states (S1 and S2). As it can be seen in \autoref{fig:forecast_results} (a), the \ac{SLP} has particularly high errors.

\begin{table}[b]
\caption{Overall \ac{RMSE} and \ac{MAPE} results in a current and future grid state.}
\label{tab:overall-rmse-results}
\centering
\footnotesize
\begin{tabular}{lrrrrrr}
	\toprule
	Estimator                           & \multicolumn{2}{c}{Current State} &
	\multicolumn{4}{c}{Future State} \\
	
	\cmidrule(lr){2-3}
	\cmidrule(lr){4-7}
	& & & \multicolumn{2}{c}{S1} & \multicolumn{2}{c}{S2} \\
	& RMSE & MAPE & RMSE & MAPE & RMSE & MAPE \\
	\cmidrule(lr){2-3}
	\cmidrule(lr){4-5}
	\cmidrule(lr){6-7}
	LSTM                       & 190.4 &  6.7\% & 352.8 & 26.2\% & 259.3 & 11.6\% \\
	CNN-LSTM                   & 182.4 &  6.2\% & 359.2 & 27.0\% & 255.5 & 12.1\% \\
	\midrule
	BM1                        &  186.8 &  6.4\% & 374.3 & 23.2\% & 276.0 & 10.6\% \\
	BM2                        &  194.2 &  6.2\% & 474.8 & 31.0\% & 341.4 & 13.5\% \\
	ARIMA                      &  246.7 &  8.0\% & 435.0 & 30.1\% & 297.0 & 11.8\% \\
	SLP                        &  579.9 & 25.0\% &  - & - & - & - \\
	\bottomrule
    \multicolumn{7}{l}{\textit{\acs{BM}1: One day ago, BM2: Day one week ago}}
\end{tabular}
\end{table}
The CNN-LSTM approach outperforms all \textit{benchmark estimators} across all scenarios using \ac{RMSE} as the comparison metric, whereas the \ac{LSTM} is superior to all other estimators in future grid states. For example in S2, the \ac{RMSE} is significantly better for the CNN-LSTM (255.5) compared to the day-ago estimator's (BM1) RMSE of 276.0 ($t(8736)=-3.46, p<0.01$). For the current state (\ac{CS}), the CNN-LSTM approach reduces the total error by 68.5\% (\ac{RMSE}) compared to the \acp{SLP}. When comparing the \textit{two \ac{ML} approaches} with each other, we find no superior approach (see the respective \ac{RMSE} and \ac{MAPE}). While the CNN-LSTM approach has a marginally lower error than the \ac{LSTM} in states with less \ac{PV} penetration (\ac{CS} and S2), the \ac{LSTM} appears to perform better in a future grid state with high \ac{PV} penetration. 
In terms of \textit{runtime performance}, the \ac{LSTM} is computationally more expensive than the CNN-LSTM. Whereas the latter only necessitates 50 training epochs (47 seconds),  training the\ac{LSTM} requires 200 epochs (1 min 13 seconds)\footnote{Instance specifications: Intel(R) Core(TM) i7-1165 CPU (clocked at 2.80 GHz), 16 GB RAM.}.
%The slightly higher training time for the \ac{LSTM} seems acceptable in the light of better predictions.

\begin{figure}
%\subfloat[Forecasting results (\ac{RMSE}) with \textbf{current} state of grid using smart meter data (case \textit{a}).\label{fig:forecast_results_current_state}]{\includegraphics[scale=0.45]{images/result_RQ_A_CurrentGrid}}
%\\
%\subfloat[Forecasting results (\ac{RMSE}) with \textbf{future} state of grid using simulated data (case \textit{b}) of Scenario S1.\label{fig:forecast_results_future_state}]{\includegraphics[scale=0.45]{images/result_RQ_B_FutureGrid_ENP}}
%\\
%\subfloat[Forecasting results (\ac{RMSE}) with \textbf{future} state of grid using simulated data (case \textit{b}) of Scenario S2.\label{fig:forecast_results_future_state_S2}]{\includegraphics[scale=0.45]{images/result_RQ_B_FutureGrid_LinFS}}
%
%\caption{Forecasting results (\ac{RMSE}) of ML models and naive estimators in a current (a) and future (b) and (c) grid state unfolded over all hours of the day. The higher errors during the day can be attributed to the uncertain actual \ac{PV} production.}
%
%
\centering
\includegraphics[width=\linewidth]{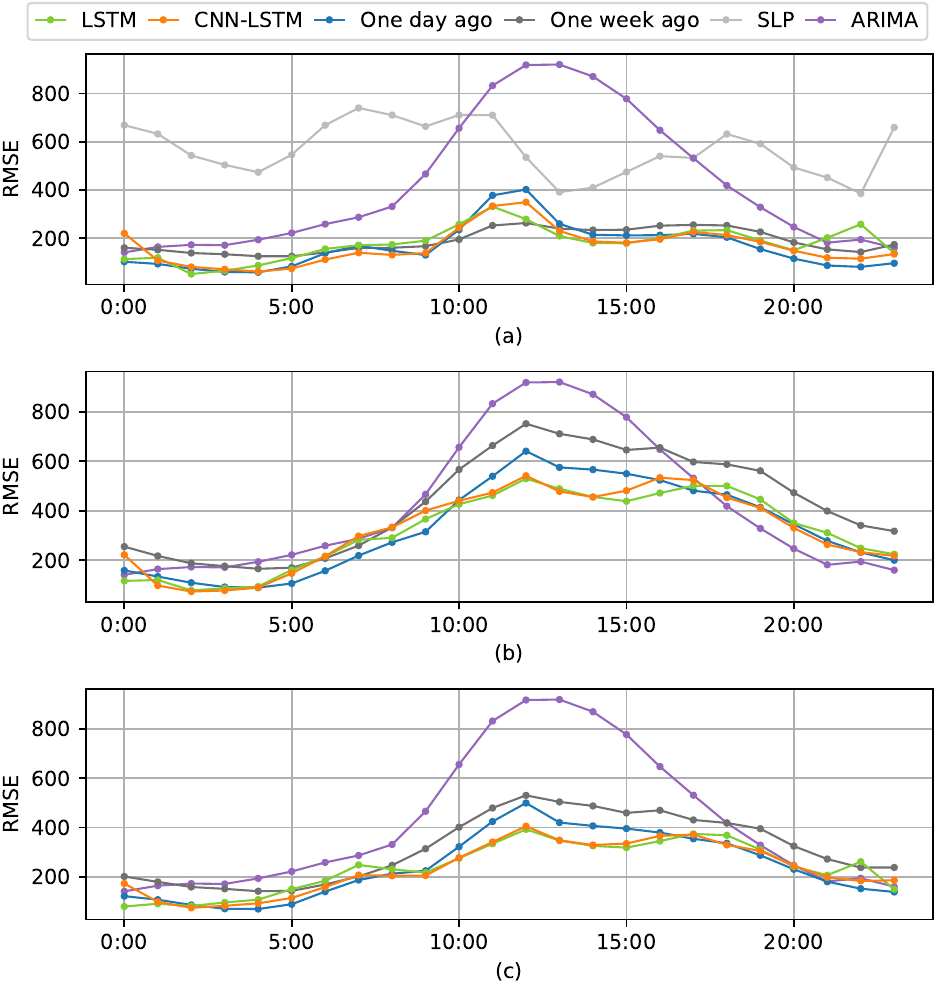}
\caption{Forecasting results (\ac{RMSE}) of ML models and naive estimators in a current (a) and future S1 (b) and S2 (c) grid state unfolded over all hours of the day. The higher errors during the day can be attributed to the uncertain actual \ac{PV} production.}
\label{fig:forecast_results}
\vspace*{-0.5em}
\end{figure}

%For the \textit{current grid state}, our results show that all \ac{ML} predictors clearly outperform \acp{SLP}. . 

%Although the naive estimators "Day before" and "Day one week before" have a comparatively low \ac{RMSE} of 30.904 and 29.954, respectively, the \ac{ML} approaches are clearly superior. Moreover, a closer look at the \ac{RMSE} at the 24-hour granularity level in figure  uncovers that the \ac{ML} approaches perform better at almost every hour of the day. Figure \autoref{fig:forecast_results_current_state} also reveals that it is more difficult for the estimators to predict electricity demand at times of day with large fluctuations in demand, such as 12 PM.

Our comparison of the demand forecasts in the \textit{current and future grid states} reveals that the performance of all estimators suffers with increased \ac{PV} generation in S1 and S2.
The higher the \ac{PV} penetration in S1 (compared to S2), the higher the prediction error for all considered models.
We posit that the primary cause of this phenomenon is the inherent unpredictability of cloudy days during summer, which the tested methodology struggles to forecast accurately. Consequently, the impact of a misprediction is higher, as the level of \ac{PV} penetration increases.
%Moreover, we see that the \ac{MAPE} has drawbacks as a metric for comparing predictive models in future grid states, as the grid demand is more often at levels near or exactly at 0.
%However, we
We can compare our results to the state of the art based on the \ac{MAPE}. In \cite{kong_short_term_2019}, the \ac{LSTM} has a \ac{MAPE} of around 8.4\%, which is slightly higher than the \ac{MAPE} of our results.

\section{Discussion}

%Although our study is in progress, we can already demonstrate that current \ac{ML} methods clearly outperform well-known demand forecasting methods used in \textit{practice} (i.e., \acp{SLP}). Hence, our study shows how important it will be in practice to rely on more effective prediction methods in the future. 
%
%When comparing \ac{ML} predictions to simple estimators, the differences are relatively small, but even small prediction errors can have large monetary impacts. We therefore recommend utilities to adopt \ac{ML}-based forecasts, which they can train based on their available data.

Our results highlight that even well-performing forecasting methods suffer from additional distributed generation and battery storage (the RMSE in all future grid states is higher than in the current state). Thus, future research needs to develop novel methods based on simulated results that provide more reliable demand forecasts for future grid states. Based on our results, we endorse the usage of digital twins for simulating future grid states.
Nevertheless, the results of TABLE~\ref{tab:overall-rmse-results} show that \ac{ML} approaches outperform the existing \acp{SLP} and also all other evaluated benchmark predictors, including \ac{ARIMA}, when using the \ac{RMSE} as a metric in the current and more importantly especially in the future grid.
% Our results also highlight the need for future research in developing methods that produce reliable demand forecasts even with much higher presence of distributed generation and \ac{BESS}.

%\subsection{Discussion on metrics}
%When comparing the results based on \ac{MAPE}, the "day ago" estimator seems to outperform the \ac{ML} approaches. A closer look at the detailed time series reveals that the demand reaches regions close to zero more often, especially the higher the \ac{PV} penetration (see also \autoref{fig:boxplot_accum_load}).
Regarding the metrics, we see that the \ac{MAPE} has drawbacks for comparing predictive models in future grid states, as the grid demand is more often at levels near or exactly zero (see Fig.~\ref{fig:boxplot_accum_load}, left plot).
%In these cases, \ac{MAPE} is an imprecise metric for evaluating prediction quality \cite{2021_Chicco_MetricsComparison}.
Hence, we recommend to focus on other evaluation metrics (e.g. \ac{RMSE}) to compare the predictive performance of models in future grid states.
%Moreover, for example, the slightly better \ac{MAPE} in S1 for the day-ago estimator (23.2\%) is not significantly better than for the CNN-LSTM (27.0\%), $t(8736)=-1.07$, p=0.28.

\section{Conclusion and future work}
Our study evaluates simple forecasting practices (e.g., \acp{SLP}) and state-of-the-art \ac{ML} methods to predict electricity demand in the current and future grid states. The results of this paper thereby challenge state-of-the-art approaches for future grid scenarios through the novel combination of a digital twin with variants of \ac{LSTM} architectures. Interestingly, we can show that higher \ac{PV} penetration rates reduce the prediction quality in future grid states regardless of the prediction method.
Considering the upcoming changes in the grid, our work demonstrates the power of \ac{ML} approaches compared to the existing baseline models for load forecasting.
Nevertheless, the increasing forecasting errors in future grid states with higher \ac{PV} penetration show the need for even more sophisticated forecasting methods.
%Considering the upcoming changes in the grid, our work highlights the need for new and more sophisticated forecasting methods to adequately support utilities in the future, and identifies areas for further research and development.
%Such a future direction can be based on \ac{ML} approaches

Our preliminary study has several limitations that we seek to address in our future research.
%In addition, we have not yet been able to investigate all aspects of our research question. 
%
%First, we will soon receive an \textit{expanded dataset} % from the cooperating utility that will cover more households and a longer time period. It will also include information on other energy systems, including heat pumps and wall boxes. This additional dataset will 
%    (i) make the predictive models more robust and potentially increase the predictive performance, 
%    (ii) extend the simulation by including additional energy systems, and 
%    (iii) allow our intended comparison between predictions on grid level and individual household level.
%We also plan to include weather data, public holidays and special events, geographic information, and household characteristics through a residential survey.
%
First, the set of currently employed \textit{predictive models} is limited. Hence, future work will comprise additional \ac{ML} approaches for demand estimation.
%So far, the study is limited to the comparison of predictive performance and training time regarding \ac{LSTM} networks and naive estimators---therefore, we plan to incorporate even more baseline models (such as linear regression, support vector regression, boosting approaches, and simple feed-forward networks) and other metrics (e.g., system cost, running time etc.) for the comparison. Also, we
More sophisticated \ac{ML} approaches will likely yield considerable predictive performance gains.  One promising approach are Transformers \cite{vaswani_attention_2017}, which already master several sequential modeling problems.
%Yet, our literature review so far has not revealed a detailed study on the application of Transformers for short-term electricity demand forecasting. We will additionally extend our parameter tuning for the \ac{ML} models.
Our future investigation will also ponder the trade-off between training time and prediction error.
Second, our \textit{digital twin} will be improved. We plan 
(i) to equip existing households with additional components (e.g., heat pumps and charging stations), and
%(ii) to evaluate the potential of sector coupling of all simulatively added components;
(ii) to reintegrate the results from this analysis in the prediction models to improve planning tasks and thus smooth peak loads on the grid level.
Third, the effect of more sophisticated  control strategies for battery storage or heating systems focusing on the general reduction of energy consumption, like \cite{2022_Bayer_Enhancing_RL_for_HVAC}, should be considered in the context of demand prediction.

\section*{Acknowledgment}
We would like to thank our research partner Stadtwerk Hassfurt GmbH for providing the dataset.
This paper is an outcome of the research project \textit{DigiSWM} (DIK0298/02) founded by the Bavarian State Ministry of Economic Affairs, Regional Development and Energy.

\bibliographystyle{IEEEtran}
\bibliography{references}

\end{document}